\documentclass[aps,prl,floatfix,twocolumn,footinbib,amsmath,amssymb]{revtex4}
\usepackage{amsmath,amsthm,amssymb,color,psfrag}
\input epsf.tex
\usepackage{latexsym}
\usepackage{graphicx}

\definecolor{darkred}{rgb}{0.6,0.0,0.0}
\definecolor{darkblue}{rgb}{0.0,0.0,0.5}
\definecolor{darkgreen}{rgb}{0.0,0.5,0.0}
\definecolor{brown}{rgb}{0.0,0.0,0.0}
\newcommand{\red}{\color{darkred}}
\newcommand{\blue}{\color{darkblue}}
\newcommand{\green}{\color{darkgreen}}

\newcommand{\Tr}{\mathrm{Tr}}
\renewcommand{\t}{t}

\begin{document}

\title{Seeing in {\blue C}{\red o}{\green l}{\blue o}{\red r}: Jet Superstructure}
\author{Jason Gallicchio}
\author{Matthew D. Schwartz}
\affiliation{Department of Physics, Harvard University,Cambridge, Massachusetts 02138, USA}

\begin{abstract}
A new class of observables is introduced which aims to characterize the superstructure
of an event, that is, features, such as color flow, which are not determined by the jet four-momenta alone.
Traditionally, an event is described as having jets which are independent objects; each jet has some
energy, size, and possible substructure such as subjets or heavy flavor content. This description discards information
connecting the jets to each other, which can be used to determine if the jets came from decay of
a color-singlet object, or if they were initiated by quarks or gluons. An example superstructure variable,
pull, is presented as a simple handle on color flow. It can be used on an event-by-event basis as a
tool for distinguishing previously irreducible backgrounds at the Tevatron and the LHC.
\end{abstract}
\maketitle

Hadron colliders, such as the LHC at CERN, are fabulous at producing quarks and gluons. At energies well
above the confinement scale of QCD, these colored objects are produced in abundance, only hadronizing into
color-neutral objects when they are sufficiently far apart. The observed
final-state hadrons collimate into jets which, at a first approximation, are in one-to-one correspondence with hard-partons
from the short-distance interaction. In fact, this description is so useful that it is usually possible
to treat jets as if they {\it are} quarks or gluons. Conversely, in a first-pass phenomenological study,
it is possible simply to simulate the production of quarks and gluons, assuming they can be accurately reconstructed
experimentally from observed jets.

In certain situations, the jet four-momenta alone do not adequately characterize the underlying hard process.
For example, when
an unstable particle with large transverse momentum decays hadronically,
the final state may contain a number of nearly collinear jets. These
jets may then be merged by the jet-finder.
Or, due to contamination from the underlying event,
 the energy of
the reconstructed jet may not optimally represent the energy of the hard parton, thereby obscuring
the short-distance event topology. Over the
last few years, a number of improved jet algorithms and filtering techniques have been developed to
improve the reconstruction of hard scattering
kinematics~\cite{Butterworth:2008iy,Kaplan:2008ie,Ellis:2009me,Krohn:2009th},
with experimentally endorsed
successes including
reviving a Higgs to $b\bar{b}$ discovery channel at the LHC~\cite{Butterworth:2008iy} (implemented by ATLAS~\cite{atlas})
and making top-tagging as reliable as $b$-tagging~\cite{Kaplan:2008ie} (implemented by CMS~\cite{Giurgiu:2009wv}).
Nevertheless, there is still a horde of information
in the events which these substructure techniques ignore. Jets have color, and are {\it color-connected} to each other,
providing the event with an observable and characterizable {\it superstructure}.

\begin{figure}[h]
\includegraphics[width=0.45\hsize]{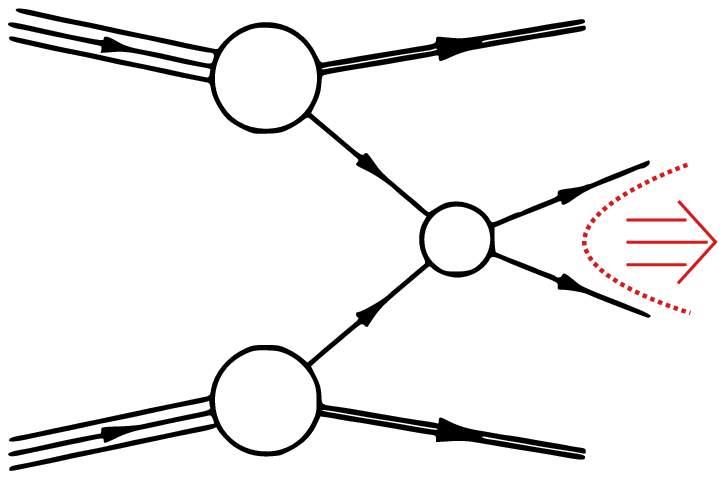}
\includegraphics[width=0.45\hsize]{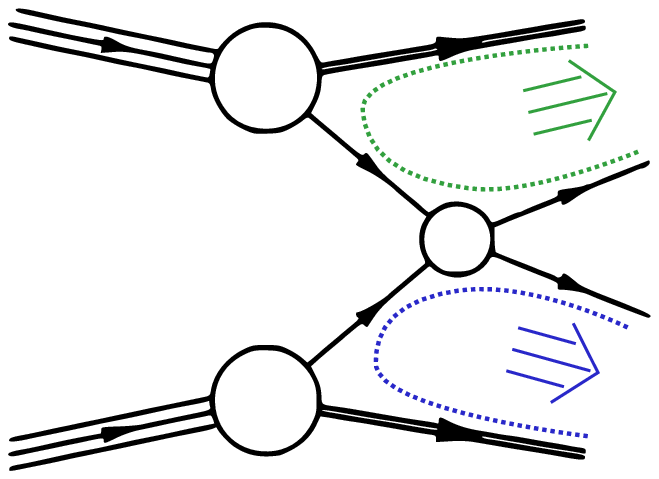}
\caption{ Possible color connections for signal ($pp\to H\to b\bar{b}$) and for background ($pp\to g \to b\bar{b}$).}
\label{fig:beam_spray_diagram}
\end{figure}

The term {\it color-connected} comes from a graphical picture of the way $SU(3)$ group indices are contracted
in QCD amplitudes.
To be concrete, consider the production of a Higgs boson at the LHC with
the Higgs decaying to bottom quarks. The hard process is $q\bar{q} \to H \to b\bar{b}$.
Since the Higgs is a color singlet, the color factor in the leading order
matrix element for this production has the form $\Tr[ T^A T^B] \Tr[T^C T^D]$, where $T^A$ are
generators of the fundamental representation of $SU(3)$, $A$ and $B$ index the initial state
quarks and $C$ and $D$ index the final-state $b$'s.
Since $\Tr[T^C T^D] \propto \delta^{CD}$, the color of $C$ must be the same as $D$,
which can be represented graphically as a line connecting quark $C$ to quark $D$. This
{\it color string} or {\it dipole} is shown in Figure~\ref{fig:beam_spray_diagram}.
An example background process is $q\bar{q} \to g\to b\bar{b}$.
Here, there are two possibilities for the color connections:
$\Tr[T^A T^C] \Tr[T^B T^D]$ and $\Tr[T^A T^D] \Tr[T^B T^C]$, both of which connect one incoming quark to one outgoing quark,
as shown also in Figure~\ref{fig:beam_spray_diagram}.
The color string picture treats gluons as bifundamentals, which is correct in the limit of a large the number of colors, $N_C \to \infty$.
Subleading corrections are included in simulations through color-reconnections, which amount to a $1/N_C^2\sim 10\%$ effect.

Since color flow is physical, it may be possible to extract the color connections of an event.
Such information would be complimentary to
the information in the jets'  four-momenta and therefore may help temper otherwise irreducible backgrounds.
For example, one application
would be in cascade decays from new physics models. In supersymmetry, one often has a large number of jets,
originating from on-shell decays like $\tilde{q} \to q \chi$ or from color-singlet
gauge boson or gaugino decays. One of the main difficulties in extracting the underlying physics from these decays is the combinatorics:
which jets come from which decay? Mapping the superstructure color connections of the events could then
greatly enhance our ability to decipher the short-distance physics.

\begin{figure}[t]
\psfrag{mpi}[B]{$-\pi$}
\psfrag{pi}{$\pi$}
\psfrag{eta}{$\mathbf{y}$}
\psfrag{phi}{$\boldsymbol{\phi}$}
\psfrag{-}[rB]{\scriptsize{$-$}}
\psfrag{0}[B]{\scriptsize{$0$}}
\psfrag{1}[B]{\scriptsize{$1$}}
\psfrag{2}[B]{\scriptsize{$2$}}
\psfrag{3}[B]{\scriptsize{$3$}}
\begin{center}
\begin{tabular}{cc}
\textbf{\ Signal}
&
\textbf{\ Background}
\\
\includegraphics[height=0.38\hsize]{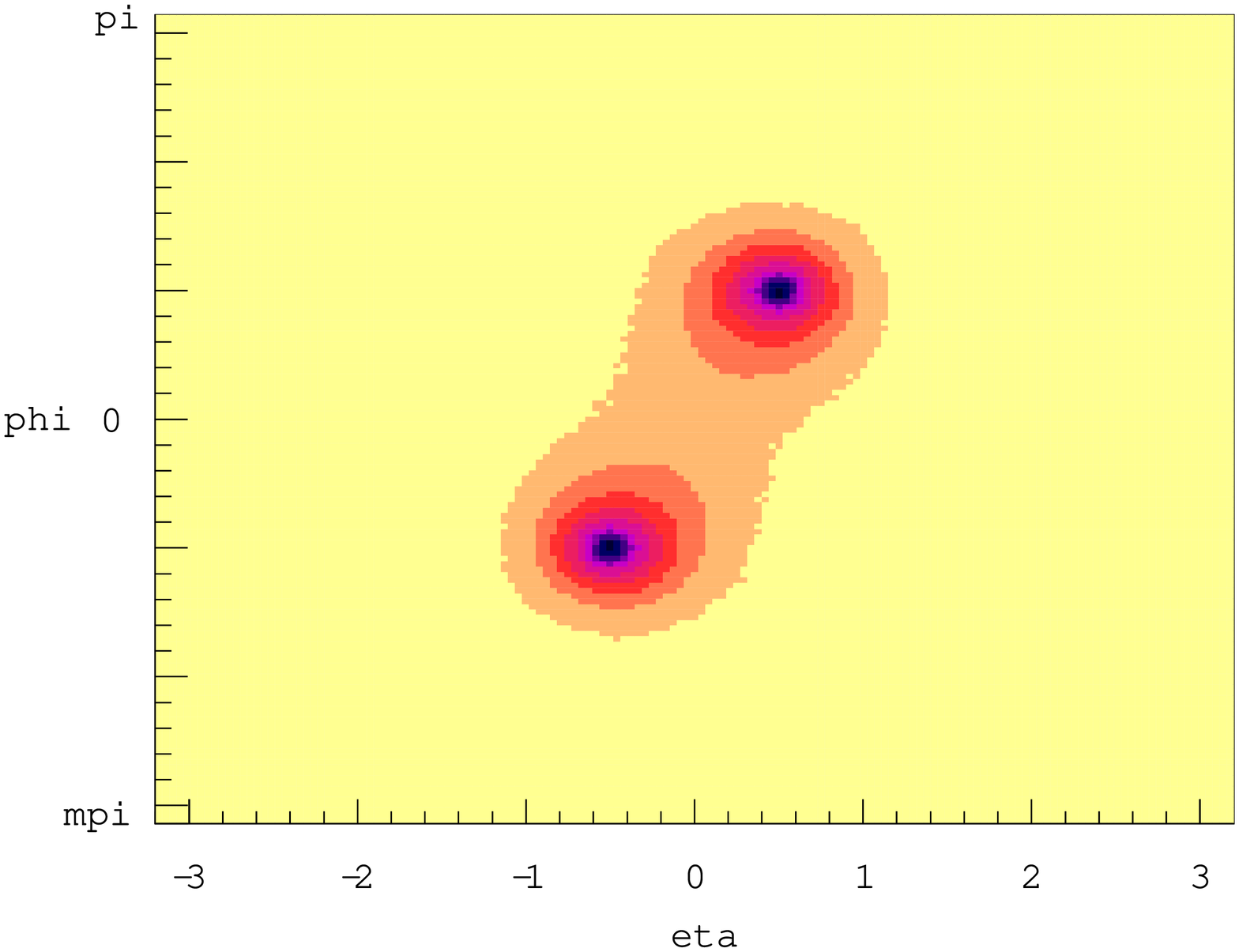}
\quad
&
\includegraphics[height=0.38\hsize]{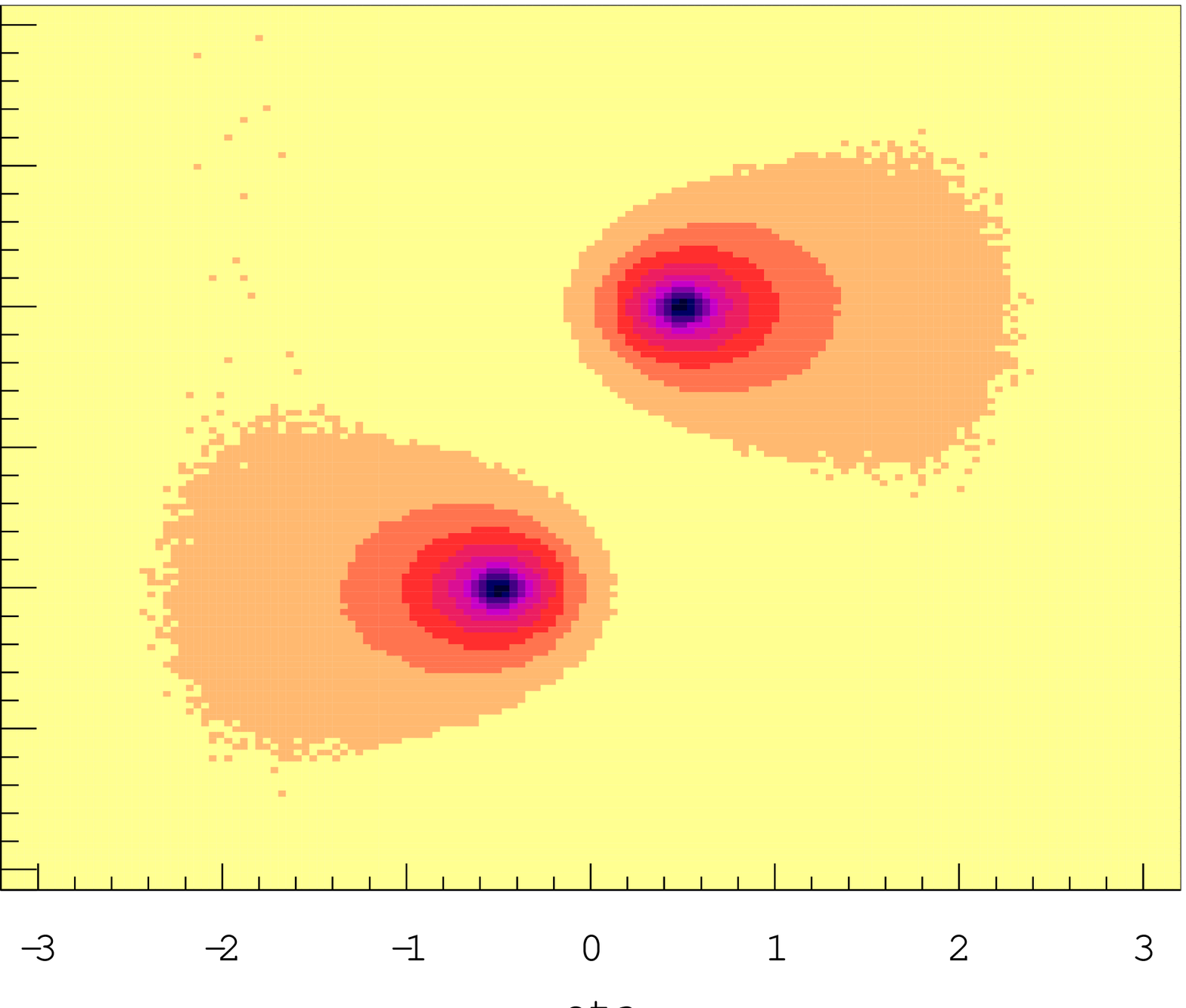}
\end{tabular}
\end{center}
\caption{Accumulated $p_T$ after showering a particular partonic phase space point 3 million times. Left
has the $b$ and $\bar{b}$ color-connected to each other (signal) and right has the $b$ and $\bar{b}$
color-connected to the beams (background). Contours represent factors of 2 increase in radiation.}
 \label{fig:acc_highres3M_raw_crop}
\end{figure}

In order to extract the color connections, they must persist into the distribution of the observable hadrons.
The basic intuition for how the color flow might show up follows from approximations used
in parton showers~\cite{Sjostrand:2006za,Skands:2009tb}.
In these simulations,
the color dipoles are allowed to radiate through Markovian evolution from the large energy scales associated with the hard interaction
to the lower energy scale associated with confinement. These emissions transpire in the rest frame of the dipole. When boosting back
to the lab frame, the radiation appears dominantly within an angular region spanned by the dipole,
as indicated by the arrows in Figure~\ref{fig:beam_spray_diagram}.
Alternatively, an angular
ordering can be enforced on the radiation (as in {\sc herwig}~\cite{Bahr:2008pv}).
The parton shower treatment of radiation attempts to include a number of features
which are physical but hard to calculate analytically,
such as overall momentum and probability conservation or coherence phenomena associated with soft radiation.

It is more important that these effects exist in data than that they are included in the simulation.
In fact, color coherence effects have already been seen by various
experiments.
In $e^+e^-$ collisions, for example, evidence for color connections between final-state 
quark and gluon jets was observed in three jet events by JADE at DESY \cite{Bartel:1983ii}. Later, at LEP,
the L3 and DELPHI experiments found evidence for color coherence
among the hadronic decay products of color-singlet objects in $W^+ W^-$ events~\cite{Achard:2003pe,Abdallah:2006uq}.
Also, in $p\bar{p}$ collisions at the Tevatron,
color connections of a jet to beam remnants have been observed by D0 in $W$+jet events~\cite{Abbott:1999cu}.
All of these studies used analysis techniques which were very dependent on the particular event topology.
What we will now show is that it is possible to come up with a very general discriminant which can
help determine the color flow of practically any event. Such a tool has the potential
for wide applicability in new physics searches at the LHC.

\begin{figure}[t]
\psfrag{th}{$\boldsymbol{\theta_\t}$}
\psfrag{mp}{$\boldsymbol{\left|\vec \t \right|}$}
\psfrag{mpi}{$-\pi $}
\psfrag{pi}{$\pi$}
\psfrag{0.04}{\scriptsize{$0.04$}}
\psfrag{0.02}{\scriptsize{$0.02$}}
\psfrag{0}{\scriptsize{$0$}}
\begin{center}
\begin{tabular}{cc}
\textbf{\quad \ Signal Pull}
&
\textbf{\quad \ Background Pull}
\\
\includegraphics[height=0.42\hsize]{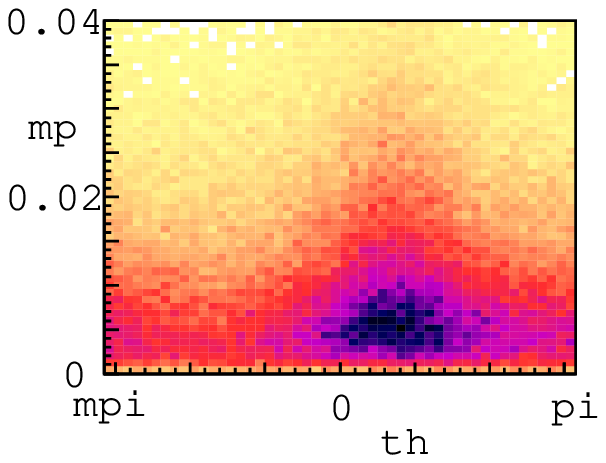}
&
\includegraphics[height=0.42\hsize]{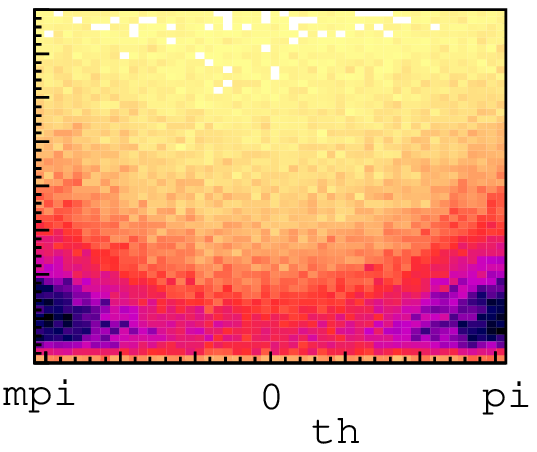}
\end{tabular}
\end{center}
\vskip -0.4cm
\caption{Event-by-event density plot of the pull vector of the $b$ jet
in polar coordinates. The signal (connected to $\bar{b}$ jet) is on the left,
the background (connected to the left-going, $y=-\infty$ beam) is on the right. 
$10^5$ events are shown.} \label{fig:twist_angles}
\end{figure}

For an example, we will use Higgs production in association with a $Z$. The $Z$ allows the Higgs
to have some $p_T$ so that its $b\bar{b}$ decay products are not back-to-back in azimuthal angle, $\phi$.
Our benchmark calculator will be {\sc madgraph}~\cite{Alwall:2007st} for the matrix elements
interfaced to {\sc pythia 8}~\cite{Sjostrand:2007gs} for the parton shower, hadronization and underlying event,
with other simulations used for validation.

To begin, we isolate the effect of the color connections by fixing the parton momentum.
We compare events with $Z b\bar{b}$ in the final state (with $Z\to$ leptons)
in which the quarks are color-connected to each other (signal)
versus color-connected to the beam (background).
In Figure~\ref{fig:acc_highres3M_raw_crop}, we show the distribution of radiation for a typical case,
where $(y,\phi)=(-0.5,-1)$ for one $b$ and $(y,\phi)=(0.5,1)$ for the other, with $p_T=200$ GeV for each $b$,
where $y$ is the rapidity.
For this figure, we have showered and hadronized
the same parton-level configuration over and over again, accumulating the $p_T$ of the final-state hadrons in $0.1\times 0.1$
bins in $y$-$\phi$ space. The color connections are unmistakable.

The superstructure feature of the jets in Figure~\ref{fig:acc_highres3M_raw_crop} that we want to isolate
is that the radiation in each signal jet tends to shower in the direction of the other jet, while in the background it
showers mostly toward the
beam. In other words, the radiation on each end of a color dipole is being pulled towards
the other end of the dipole. This should therefore show up in a dipole-type moment constructed from the radiation in or
around the individual jets.
 For dijet events, like those shown in Figure~\ref{fig:acc_highres3M_raw_crop},
one could imagine constructing a global event shape from which the moment could be extracted. However,
a local observable, constructed only out of particles within the jet, has a number of immediate advantages.
For one, it will be a more general-purpose tool, applying to events with any number of jets. It
should also be easier to calibrate on data, since jets are generally better understood experimentally than
global event topologies. Therefore, as a first attempt at a useful superstructure variable, we
construct an observable out of only the particles within the jets themselves.

In constructing a jet moment, there are a number of ways to weight the momentum, such as by  energy or $p_T$, and
to define the center the jet. These are all basically the same, but we have found
that the most effective combination is a $p_T$-weighted vector, which we call \emph{pull}, defined by
\begin{equation}
\vec \t = \sum_{i \in \text{jet}} \frac{ p_T^i \, |r_i|  }{ p_T^{\text{jet}} }\, \vec r_i \label{centeq} \, .
\end{equation}
Here, $\vec r_i = (\Delta y_i,\Delta \phi_i) = \vec c_i - \vec J$, where $\vec J= (y_J,\phi_J)$
is the location of the jet and $\vec c_i$ is the position of a
cell or particle with transverse momentum $p_T^i$. Note that we use rapidity $y_J$ for the jet instead
of pseudorapidity ($\eta_J$); because the jet is massive this makes $\vec r_i$
boost invariant and a better discriminant (rapidity
and pseudorapidity are equivalent for the effectively massless cells/particles, $\vec c_i$).
The centroid (Eq.~\eqref{centeq} without the $|r_i|$ factor) is usually almost identical to $\vec J$,
 the location of the jet four-vector in the $E$-scheme (the sum of four-momenta of the jet constituents).

An important feature of the pull vector $\vec \t$ is
that it is infrared safe. If a very soft particle is added to the jet, it has negligible $p_T$,
and therefore a negligible effect on $\vec \t$. Moreover, since
pull is linear in $p_T$, if a particle splits into two collinear particles at the same $\vec r$, the pull
vector is also unchanged. This property guarantees that pull should be fairly insensitive to fine details
of the implementation, such as the spatial granularity or energy resolution of the calorimeters.

\begin{figure}[t]
\psfrag{th}{$\boldsymbol{\theta_\t}$}
\psfrag{signal}{ signal}
\psfrag{background}{ \blue background}
\psfrag{towardOtherJet}{ toward other jet}
\psfrag{PYTHIA6}{ {\sc pythia}6}
\psfrag{PYTHIA8}{ {\sc pythia}8}
\psfrag{HERWIGpp}{ {\sc herwig}++}
\psfrag{mpi}{$-\pi $}
\psfrag{pi}{$\pi$}
\psfrag{0}{$0$}
\psfrag{1}{$1$}
\psfrag{2}{$2$}
\psfrag{3}{$3$}
\includegraphics[width=0.75\hsize]{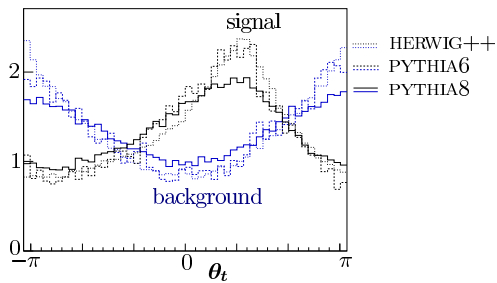}
\caption{
Distribution of the pull angle (for the $b$ jet) with
$\Delta y_{b\bar{b}} = 1$ and $\Delta \phi_{b\bar{b}} = 2$,
for signal and background, showered $10^5$ times with different
Monte Carlos.
} \label{fig:distdisc}
\end{figure}

The event-by-event distribution of the pull for the left $b$ jet from Figure~\ref{fig:acc_highres3M_raw_crop}
is shown in Figure~\ref{fig:twist_angles} in polar coordinates,
$\vec \t = (|\vec \t| \cos \theta_t,| \vec \t| \sin \theta_t)$, where $\theta_\t= 0$ points towards the right-going beam,
 $\theta_t= \pm \pi$ points towards the left-going beam,
and $\theta_t \approx 0.7$ toward the other $b$ jet.
This figure shows density plots of the $\vec \t$
distributions on an event-by-event basis for the signal and background cases for this particular fixed parton-level
phase space point.
For this figure, we use as input the four-momenta of all long-lived observable particles. If instead, we use the hadronic energy
in $0.1\times 0.1$ cells treated as massless four-vectors, the distribution of pull vectors is nearly identical.

We can see that most of the discriminating information is in the pull angle, $\theta_\t$, rather than the magnitude $| \vec \t|$.
This leads to Figure~\ref{fig:distdisc}, which shows the distribution of the pull angle for the signal
and the background in this particular kinematic configuration. This figure also shows that the pull vector is
not particularly sensitive to the Monte Carlo program used to generate the sample; the pull angle
distributions for {\sc herwig++ 2.4.2}~\cite{Bahr:2008pv}, {\sc pythia 8.130}~\cite{Sjostrand:2007gs},
and {\sc pythia 6.420} with the $p_T$-ordered shower~\cite{Sjostrand:2006za} are all quite similar.

The previous three figures all have the parton momentum fixed. Similar distributions result from other phase space points.
We fixed the parton momentum to show the usefulness of pull in situations
which would be indistinguishable using the jet four-momenta alone.
This exercise controls for correlations between pull and matrix-element-level kinematic discriminants.
Also, note that there is another possible color-flow for the background events,
where the left-going jet is color-connected to the right-going beam. Then,
the most-likely pull angle would be more similar to the signal.
Fortunately, this only occurs about 10\% of the time for the dominant background.

\begin{figure}[t]
\psfrag{th}[l]{$\boldsymbol{\Delta \theta_\t}$}
\psfrag{signal}[l]{ signal}
\psfrag{background}[l]{\blue background}
\psfrag{mpi}[l]{$-\pi$}
\psfrag{pi}[l]{$\pi$}
\psfrag{0}[l]{$0$}
\psfrag{1}[l]{$1$}
\psfrag{2}[l]{$2$}
\psfrag{3}[l]{$3$}
\begin{tabular}{cc}
\textbf{\quad Pull of Higher $\mathbf{p_T}$ $\mathbf{b}$}
&
\textbf{\quad Pull of Lower $\mathbf{p_T}$ $\mathbf{b}$}
\\
\includegraphics[width=0.45\hsize]{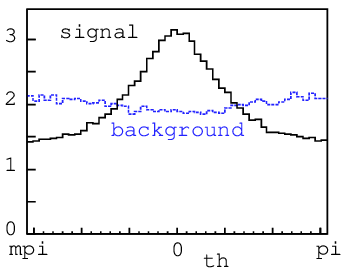}
&
\includegraphics[width=0.45\hsize]{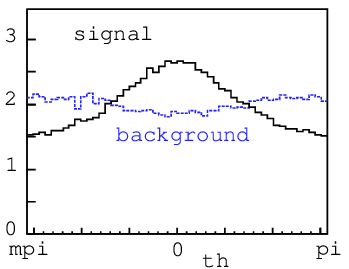}
\end{tabular}
\caption{Pull angles in the $b$ or $\bar{b}$ jet in $HZ\to Z b\bar{b}$ signal events and their $Z+b \bar b$ backgrounds.
For each event, $\Delta \theta_\t=0$ is defined to point toward the other $b$ jet. $3\times10^5$ events are shown.}
\label{fig:angle_to_other_jet}
\end{figure}

The next step is to see if pull is useful given the full distribution of signal and background events at the LHC.
The pull angle for the full $ZH\to Z b\bar{b}$ signal and $Zb\bar{b}$ backgrounds still presents a strong discriminant,
as can be seen in Figure~\ref{fig:angle_to_other_jet}.
Here, we have performed a full simulation with
{\sc madgraph 4.4.26}~\cite{Alwall:2007st} and {\sc pythia 8.130}~\cite{Sjostrand:2007gs},
including underlying event and hadronization. We choose a parton-level cut of $p_T > 15$ GeV for the $b$ quarks,
find the jets with the anti-$k_T$ algorithm with $R=0.7$, require the reconstructed mass to be
within a 20 GeV window around the Higgs mass (120 GeV), and construct the pull
angle on the radiation within each jet.

Next, let us consider some other possibilities. It is natural to look at higher moments, such as those
contained in the covariance tensor
\begin{equation}
{\mathbf C} =
\sum_{i \in \text{jet}} \frac{ p_T^i |r_i|}{ p_T^{\text{jet}} }
\left(
\begin{array}{cc}
  \Delta y_i ^2 & \Delta y_i \, \Delta \phi_i \\
  \Delta \phi_i \, \Delta y_i & \Delta \phi_i ^2
\end{array}
\right) \, .
\end{equation}
The eigenvalues $a \ge b$ of this tensor are similar to the semimajor and semiminor axes of an elliptical jet.
The overall size of these $g=\sqrt{a^2+b^2}$ provides a decent characterization of whether
the jet is initiated by a quark or gluon. Gluon jets, since they cap two color dipoles, generally have more radiation
and lead to jets with larger values of $g$. However, $g$ is strongly correlated with the mass of a jet and
the mass-to-$p_T$ ratio. Since mass and $p_T$ are contained in the jet four-momentum, this measure of
size is not likely to provide a new handle for irreducible backgrounds.
%
%
Other combinations of second-moment eigenvalues, such as the eccentricity $e=\sqrt{(a^2-b^2)/a}$ or orientation
of the ellipse, seem
much less useful. While one might expect gluon jets to be fairly elliptical, due to their being
pulled in two directions, in fact quarks turn out to be equally elliptical;
we have not found a significant difference in the eccentricity of quark and gluon jets.
Going to third or higher moments is straightforward, but serves no immediate purpose.

We conclude that the pull angle is the most useful moment-type observable for determining the color superstructure of
an event. Besides moments, one could attempt to use  more global observables, such as the amount of radiation around or between jets.
As we have mentioned, such an approach is in principle promising, but the analysis would have to be very
 process-dependent. A nice feature of pull is its universality.
Although we have used as a canonical example Higgs production in association with a $Z$ boson,
the pull angle can be used to characterize any process with jets, such as cascade decays in
supersymmetry or resonance decays in composite models.
In fact, for practically any new physics scenario involving jets,
finding the color connections would be very helpful, and the pull angle provides a simple tool to extract this information.

In order to apply superstructure variables to new physics searches, it will be critical to first validate them on
standard model data. One useful class of events is $t\bar{t}$ production. For semileptonic $t\bar{t}$ decays,
we can get an arbitrarily clean sample by tightening the $b$-tags, top mass window, and leptonic $W$ reconstruction. This
will give us a pure sample of hadronic, boosted $W$ bosons. The two light quark jets from the $W$ decay should be color-connected,
and the pull angle of each quark can be measured on data.
 The same sample also provides $b$ jets connected to the beam.
We have tested this idea in simulations of $t\bar{t}$ events, and have found that the pull angle distribution in
the hadronic $W$ decay products is in fact similar to that of the the Higgs decay in Figure~\ref{fig:angle_to_other_jet}.

Finally, let us mention a few words
about the choice of jet algorithm. Using the program {\sc fastjet v2.4}~\cite{Cacciari:2005hq} for
jet finding, we found that the anti-$k_T$\cite{Cacciari:2008gp} algorithm,
which takes radiation from more circular regions, gives better results than $k_T$~\cite{Catani:1993hr},
SIScone~\cite{Salam:2007xv}, or Cambridge/Aachen~\cite{Dokshitzer:1997in}.
It is also possible to find the jets with one algorithm and size,
say $R=0.7$ and then use a larger size, say $R=1.2$, to calculate the moment. We have
not found an obvious improvement from doing this, but such possibilities should be explored. For example,
if the pull angle were to be used by an experimental collaboration in Higgs search, a few percent improvement
could probably be gained by  optimizing the algorithm in coordination with the detailed experimental parameters.
It would also be worth investigating whether jet filtering~\cite{Butterworth:2008iy}
or trimming~\cite{Krohn:2009th},
could help make
pull or other superstructure variables even more discriminating.
Although there is still a lot of room for improvement, it is clear that color flows and jet superstructure
can be useful observables at hadron colliders, and are worth understanding better.

The authors would like to thank David E. Kaplan, David Krohn, Steve Mrenna,  Michael Peskin and Keith Rehermann
for useful discussions and the Jet Substructure Workshop at the University of Washington,
the FAS Research Computing Group at Harvard University
and the Department of Energy, under Grant DE-AC02-76CH03000, for support.

\end{document}